\RequirePackage{ifpdf}
\documentclass{PoS_oldRef}

\title{Enhancing Science from Future Space Missions and Planetary Radar with the SKA}

\ShortTitle{Spacecraft Tracking with SKA}

\author{{Dayton L.~Jones}$^1$, Joseph Lazio$^1$
\\ 
$^1$Jet Propulsion Laboratory, California Institute of Technology
\\
E-mail: \email{dayton.jones at jpl.nasa.gov}
}

\abstract{Both Phase 1 of the Square Kilometre Array (SKA1) and the full SKA
have the potential to dramatically increase the science return from future
astrophysics, heliophysics, and especially planetary missions, primarily
due to the greater sensitivity ($\rm{A}_{EFF}$ / $\rm{T}_{SYS}$) compared
with existing or planned spacecraft tracking facilities.  While this is not
traditional radio astronomy, it is an opportunity for productive synergy
between the large investment in the SKA and the even larger investments in
space missions to maximize the total scientific value returned to society.
Specific applications include short-term increases in downlink data rate
during critical mission phases or spacecraft emergencies, enabling new
mission concepts based on small probes with low power and small antennas,
high precision angular tracking via VLBI phase referencing using in-beam
calibrators, and greater range and signal/noise ratio for bi-static
planetary radar observations.  Future use of higher frequencies (e.g., 32
GHz and optical) for spacecraft communications will not eliminate the need
for high sensitivities at lower frequencies.  Many atmospheric probes and
any spacecraft using low gain antennas require frequencies below a few GHz. 

The SKA1 baseline design covers VHF/UHF frequencies appropriate for some
planetary atmospheric probes (band 1) as well as the standard 2.3 GHz deep
space downlink frequency allocation (band 3).  SKA1-MID also covers the
most widely used deep space downlink allocation at 8.4 GHz (band 5).  Even
a 50\% deployment of SKA1-MID will still result in a factor of several
increase in sensitivity compared to the current 70-m Deep Space Network
tracking antennas, along with an advantageous geographic location.  The
assumptions of a 10X increase in sensitivity and 20X increase in angular
resolution for SKA result in a truly unique and spectacular future
spacecraft tracking capability.}

\FullConference{
Advancing Astrophysics with the Square Kilometre Array\\
June 8-13, 2014\\
Giardini Naxos, Sicily, Italy}

\newcommand{\skipthis}[1]{}

\begin{document}
% If the speaker is not the first author, then enter the first author
% information below, so it appears in the header of each page
% this must go AFTER the \begin{document} command
\makeatletter
\setbox\@firstaubox\hbox{\small Dayton Jones}
\makeatother

\section{Introduction}

Ever since the dramatic increase in sensitivity afforded by the SKA was widely recognized, it has been clear that, in addition to the multitude of astronomical applications described in this book, such capability could provide valuable benefits in other fields as well.  One example is enhancing the science return from planetary exploration missions.  The sensitivity of SKA will greatly exceed that of any existing or planned dedicated spacecraft tracking facility.  This suggests the possibility of inter-agency collaborations that help both the radio astronomy and space science communities to make the most productive use of their major investments. 

\section{Telemetry Reception}

\begin{figure}[!hb]
\hskip 45pt \includegraphics[angle=-0,scale=0.43]{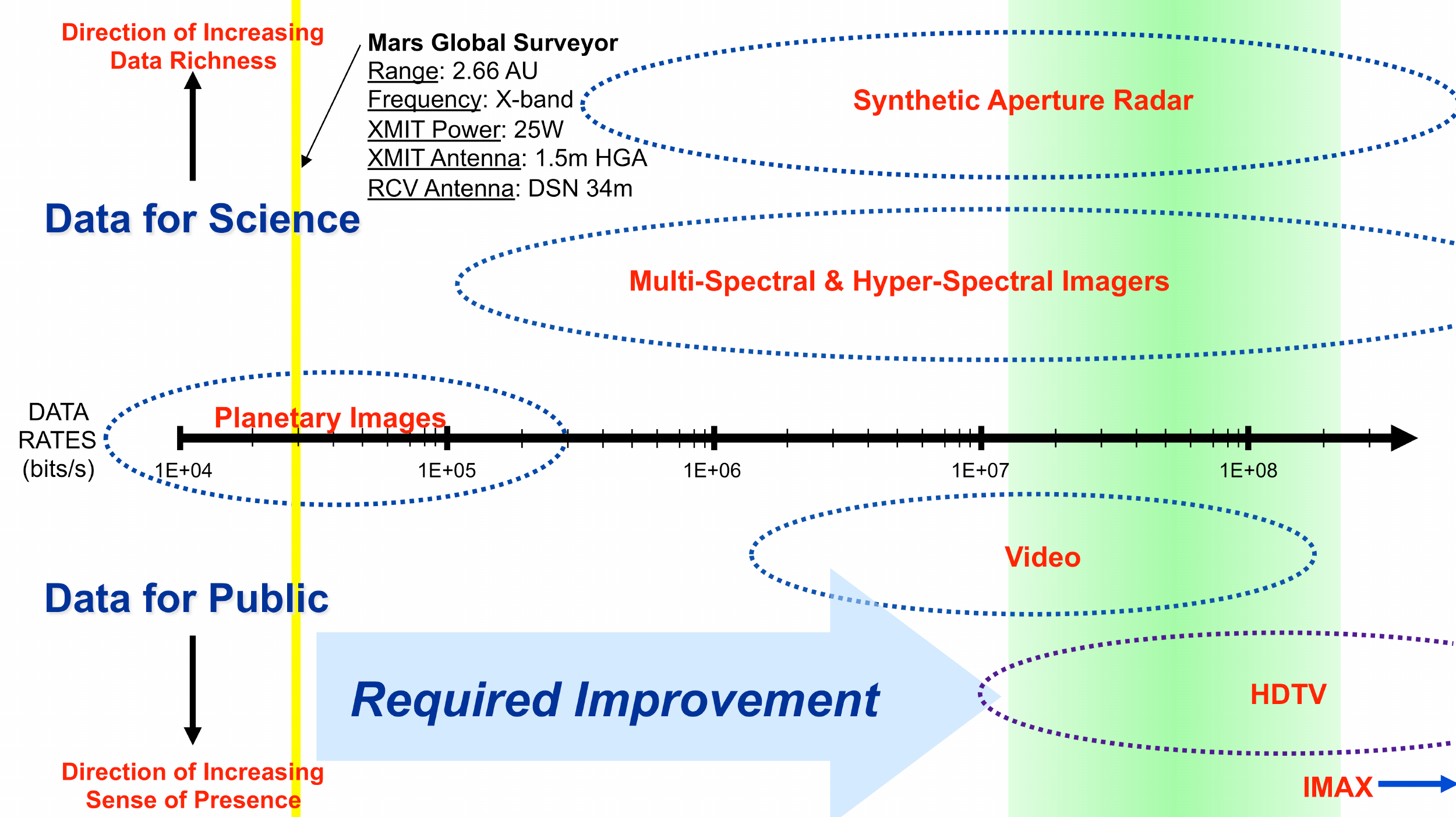}
\caption{Different types of observation possible with different data rates.  The Voyager data rates were $\sim 10$ kb/s, while observing other planets as we do Earth requires data rates $> 10$ Mb/s.  \label{fig1}}
\end{figure}

The most obvious use of increased sensitivity in the context of scientific space missions is a potentially large increase in downlink data rate, particularly from distant spacecraft. Previous examples of this include the use of Parkes and the VLA to receive telemetry during the Voyager flybys of Uranus and Neptune (Brown et al.~1986; Ulvestad et al.~1988; Brown et al.~1990).  This can be utilized to support faster data return during short-duration, high priority events or to increase the total data return through periodic tracking of longer duration missions.  In both cases, higher ground antenna sensitivity will help reduce the mismatch between the very high data rates produced by current and future flight instruments and the more limited data rates that can be supported by dedicated spacecraft tracking antennas.  This generic problem, and the importance of downlink data rate in terms in observational capabilities, is illustrated in Figure \ref{fig1}.  It should be noted that investments to improve the sensitivity of telemetry reception antennas on the ground benefit all future space missions, while investments to improve the transmitter power or antenna size on a spacecraft benefit only a specific mission.  Thus, maximizing the available sensitivity of ground tracking antennas through shared use of SKA is the more cost-effective approach to increasing the total science return from future missions, and thereby maximizing the scientific value of these expensive programs.

\subsection{Small Probes}

Small probes of planetary atmospheres and surfaces tend to have low power, low-gain antennas, and often very limited lifetimes.  A few specific examples are Venus surface probes, probes of the inner solar corona, icy moon surface probes, clusters of probes descending into gas giant atmospheres, networks of small surface probes, atmospheric balloons/gliders/aircraft, boats on Titan lakes, and probes in extreme radiation environments.  For most of these concepts optical communication is not an option because accurate pointing is not possible or because of atmospheric absorption.  Consequently it can be challenging to obtain the desired quantity of data from such probes.  

A common feature of all mission concepts of this type is the need to get data from the probes as rapidly as possible, before they die.  In the past this has been done by transmitting data from a probe to a relatively nearby (and much larger) orbiting or flyby spacecraft, which then relays the data to Earth over a high-bandwidth downlink.  This operational approach has been done successfully several times in the past (e.g., the Galileo and Cassini/Huygens missions), but always at the cost of increased mission risk and complexity.  In addition, the fuel required to put a data relay spacecraft into orbit about the target solar system body can be a significant fraction of the total spacecraft mass, and thus a significant cost driver.  

\begin{figure}[b]
\hskip 50pt \includegraphics[angle=-0,scale=0.43]{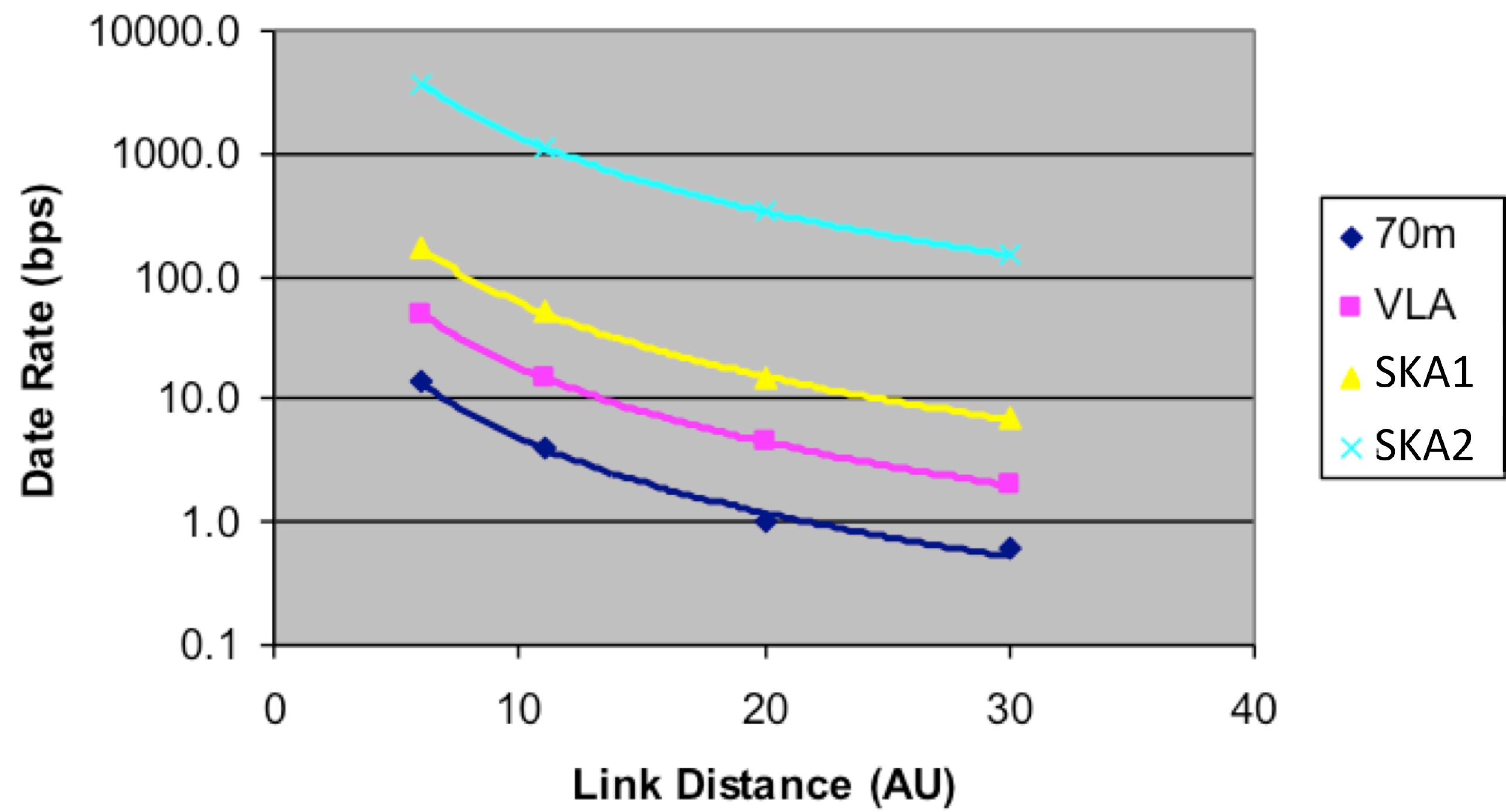}
\caption{Predicted data rates to Earth as a function of distance for several ground arrays.  These data rates are for a Huygens-like probe with a 25 W transmitter and 4 dBi gain antenna.  \label{fig2}}
\end{figure}

The SKA will allow some future small probe missions to downlink data directly to Earth, avoiding the need for a local relay spacecraft.  As an example, Figure \ref{fig2} shows the direct-to-Earth data rates that could be obtained from a probe with a 25 W transmitter (similar to the Huygens probe to the surface of Titan) and a low gain (4 dBi) antenna.  The Huygens probe transmitted data at 8 kb/s, but only to the nearby Cassini orbiter.  Huygens telemetry could not be received directly on Earth, although the carrier signal was detected.  Figure \ref{fig2} incorporates some optimistic assumptions, but still illustrates the large improvement possible in the SKA era.  This improvement could enable entirely new classes of missions involving large numbers of small probes transmitting simultaneously. 

\begin{figure}[!h]
\hskip -15pt \includegraphics[angle=0,scale=0.72]{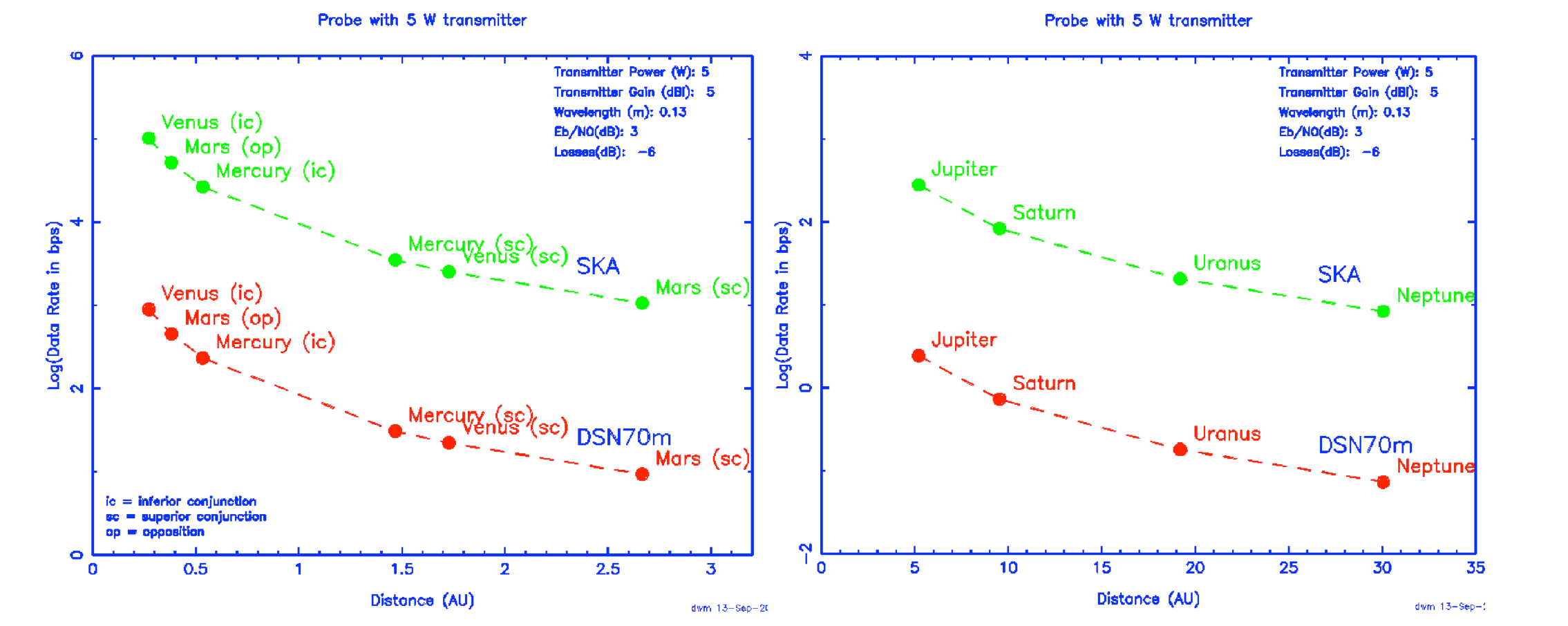}
\caption{Data rates to Earth as a function of distance for the inner planets (left panel) and outer planets (right panel).  In the left panel, ic = inferior conjunction, sc = superior conjunction, and op = opposition, the minimum and maximum distances from Earth for each planet.  The curves in both panels assume a 5 W transmitter and low gain antenna at 2.3 GHz. \label{fig3}}
\end{figure}

A somewhat more extreme example is show in Figure \ref{fig3}, which is based on a probe with a very low power (5 W) transmitter.  Even in this case the SKA allows small but useful data rates to Earth even from the outer planets.

Table 1 shows similar information in a more quantitative format.  The differences in data rates are large enough to make direct-to-Earth downlink for small probe missions viable in many cases where they would not be viable without use of the SKA. 

\begin{center}
\begin{tabular}{|c|c|c|}
\hline
\bf Telemetry from  & \bf DSN 70-m & \bf SKA \\
\hline
Venus (0.3 AU) & 11 kb/s & 1.3 Mb/s \\
\hline
Venus (2.4 AU) & 180 b/s & 21 kb/s \\
\hline
Mars (0.6 AU) & 2.9 kb/s & 340 kb/s \\
\hline
Mars (2.6 AU) & 150 b/s & 17 kb/s \\
\hline
Jupiter & 35 b/s & 4.1 kb/s \\
\hline
Saturn & 10 b/s & 1.1 kb/s \\
\hline
Uranus & 3 b/s & 330 b/s \\
\hline
Neptune & 1 b/s & 130 b/s \\
\hline
\end{tabular}
\end{center}
\vskip 12pt

Of course larger spacecraft could benefit in the same way, especially those at extreme distances from Earth such as future interstellar probe missions (e.g., Wallace et al. 2000; McNutt et al. 2002).  

\subsection{Spacecraft Emergencies}

One of the standard components of a safe-mode response to unexpected events or errors on spacecraft is to switch to a low-gain antenna for Earth communication.  This procedure is used because, in an emergency, one cannot assume that the spacecraft will retain the ability to point a high gain antenna at Earth.  Switching from a high gain antenna to a low gain spacecraft antenna can reduce the received signal level by up to several orders of magnitude.  In order to compensate for this the transmitted data rate must be reduced to very low levels, often only a few hundred or even tens of bits per second.  But during a spacecraft emergency there is a great need for engineering data as rapidly and with as much detail as possible to allow problems to be diagnosed before the situation becomes unsalvageable.  Use of the SKA, with its unequaled sensitivity, would substantially improve the chances of recovering at least one expensive mission during the SKA's planned lifetime.  

\subsection{Observing Duty Cycle}

The SKA will be built as an astronomical research facility, and its use for spacecraft tracking support will necessarily need to be limited.  Many situations that most benefit from higher ground antenna sensitivity are intrinsically of short duration, and can usually be scheduled to allow visibility from a particular site on Earth.  Such situations include planetary atmospheric entry and landing phases, short-lived small probes, occultations of spacecraft signals, or bi-static planetary radar experiments.  Consequently an operational scenario involving infrequent but high priority observations should be viable.  

\section{Radio Science}

Precise tracking of the characteristics of signals from spacecraft has enabled study of the properties of solar system bodies, including their rings, atmospheres, and interiors; has advanced our understanding of the solar wind; and has provided limits on gravitational waves at frequencies complementary to those that will be probed by the SKA Pulsar Timing Array (Asmar et al.~2005; Tellmann et al.~2009).  In many cases, particularly occultations and other dynamic events near solar system bodies or in studying the solar wind, increasing the integration time to obtain a high signal-to-noise ratio is not an option.  Only high instantaneous sensitivity can provide the requisite signal-to-noise ratio.  For short-duration radio science measurements, SKA1-MID would enhance, or even enable, observations through its increase in A$_{\rm{EFF}}$~/~T$_{\rm{SYS}}$ over current spacecraft tracking facilities. 

\section{Tracking and Navigation}

Interplanetary spacecraft navigation is an essential component of all planetary science missions.  The accuracy with which spacecraft trajectories can be predicted determines the viability of some mission concepts, and can have a large influence on the mass of fuel that must be carried.  Spacecraft tracking and navigation relies on an accurate planetary ephemeris, and three types of observational data:  Doppler (one-way or multi-way), range, and very long baseline interferometry (VLBI).  Together these provide the three position and three velocity values that define spacecraft motion.  Range provides the radial (line-of-sight) component of position, Doppler the radial component of velocity, and VLBI the two plane-of-sky components of position and velocity.  In general Doppler and range measurements can be made with narrow bandwidths, and are less likely to need the higher sensitivity offered by SKA.  VLBI, on the other hand, is often sensitivity limited and would greatly benefit from use of the SKA. 

Most VLBI spacecraft tracking is currently done with single baselines using group delays (delta differential one-way ranging or $\Delta$DOR in the spacecraft tracking community).  Narrow signals widely separated in frequency are broadcast from the spacecraft, and the observed phase slope between these signals gives the group delay on the baseline between antennas.  This works well for single baselines because phase cycle ambiguities are much easier to resolve for group delays, but the phase delay is an intrinsically more accurate observable.  A multi-baseline VLBI array using phase referencing to an angularly nearby position calibration source can measure spacecraft positions with higher accuracy (e.g., Jones et al. 2011).

\begin{figure}[!h]
\hskip 18pt \includegraphics[angle=-90,scale=0.60]{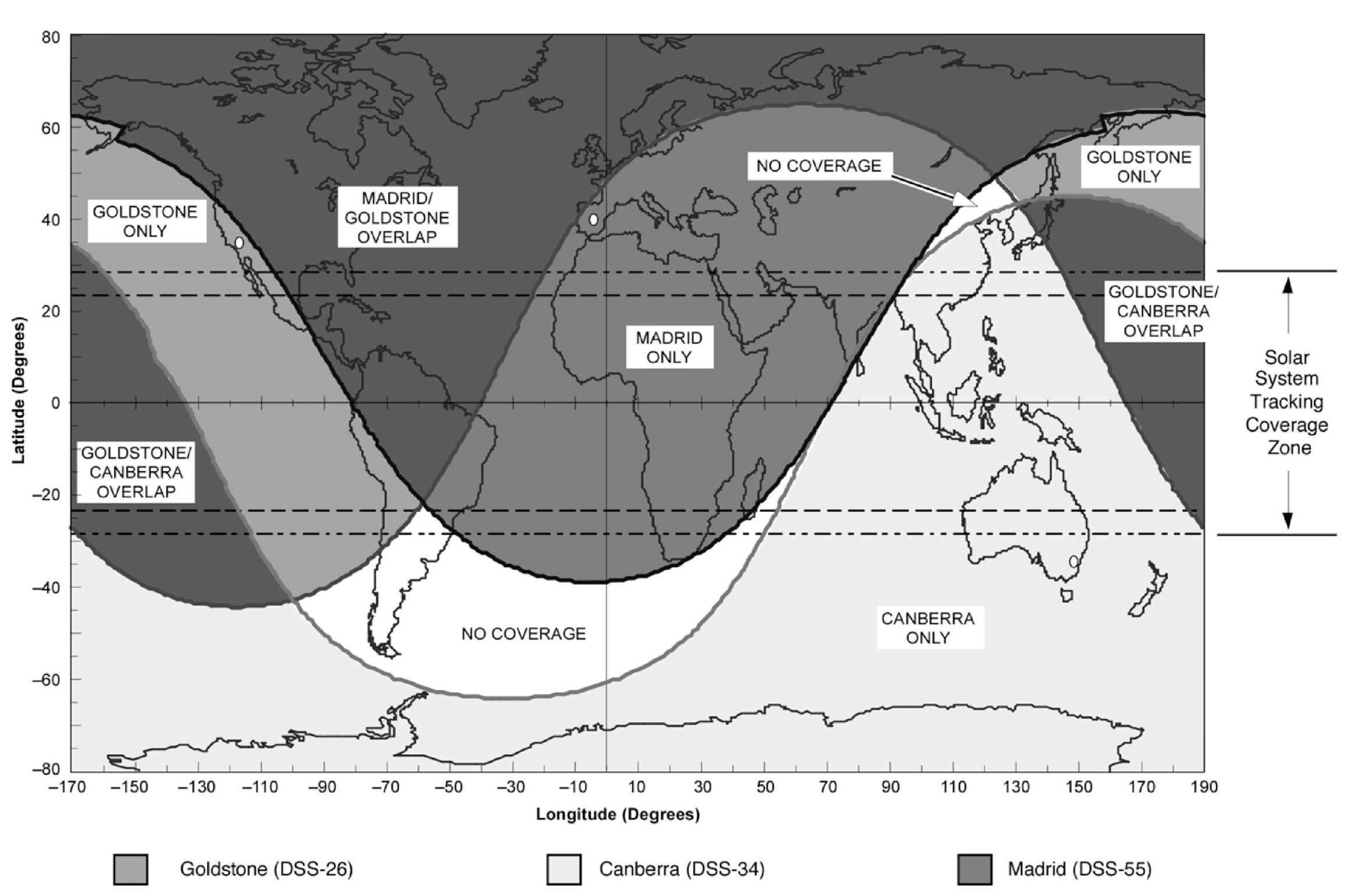}
\caption{The darkest shaded areas show where VLBI coverage is possible between DSN sites.  Lighter shaded areas show where there is visibility from only a single DSN site and VLBI tracking by the DSN is not possible.  Most deep space tracking occurs within 30 degrees of the equator (Sniffin 2012).  \label{fig4}}
\end{figure}

Neither SKA1-MID nor SKA1-SUR provide long baselines for VLBI, but it should be possible to use the long and sensitive baseline between South Africa and Australia for astrometry at overlapping frequencies (see Figure \ref{fig4}).  For astrometry of spacecraft, the overlapping frequency range must include band 3 (see section 6 for more details on relevant frequency bands).  A sufficiently sensitive single baseline can measure phase delays with respect to a compact calibration source if the angular distance between the calibration source and the spacecraft is close enough that the phase ambiguities can be resolved (e.g., Lanyi 2011), and the baseline between South Africa and Australia arrays should be sensitive enough to allow this.  The large field of view provided by small antenna diameters helps greatly here because it increases the probability of having a suitable phase calibration source within the field of view.  Having an in-beam calibrator avoids errors associated time interpolation of phases between two pointing positions, and also allows VLBI to occur simultaneously with telemetry reception.  The still larger field of view of SKA1-SUR could be combined with sub-arraying of the SKA1-MID antennas.

Figure \ref{fig4} shows the regions of mutual sky visibility between the three sites of the NASA Deep Space Network (DSN).  Note the large area around Australia and Africa where there is no VLBI coverage possible by the DSN, only visibility from a single DSN site or from no sites.  Recently cooperative observations between the DSN and a European Space Agency tracking antenna in Argentina have filled in some of the VLBI coverage gaps, but with reduced sensitivity.  SKA1-MID and SKA1-SUR would completely fill the gaps in coverage while providing higher sensitivity.  The full SKA, of course, will provide multiple long, very high sensitivity baselines and consequently it will be possible to obtain high precision astrometric positions for spacecraft in near-real time.  Most importantly, SKA will have calibration sources within its field of view at all times.

\section{Planetary Radar}

The signal-to-noise ratio for radar observations decreases as distance to the fourth power, and with radar cross-section.  A large increase in antenna sensitivity would increase the range and decrease the size of solar system object that could be detected, enhance the quality of polarization measurements and orbital solutions, and improve the resolution of radar imaging (Butler et al. 2004; De Pater 1999).

\subsection{Range and Orbit Determination}

\begin{figure}[!hb]
\hskip 40pt \includegraphics[angle=-90,scale=0.42]{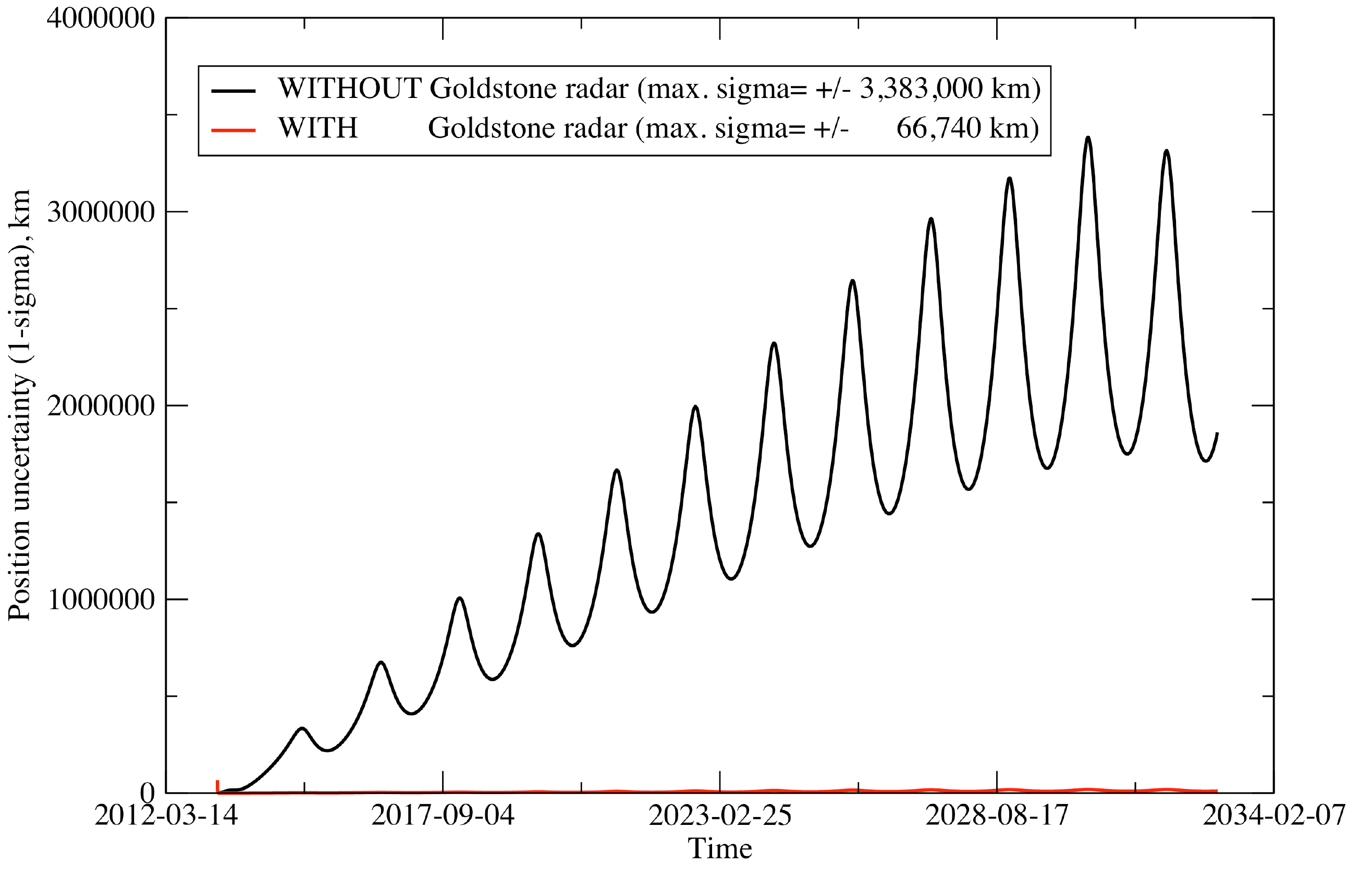}
\caption{Example of asteroid orbit uncertainty (Asteroid 2013 FB8, gravity dynamics only).  Note that the red line showing the uncertainty after a single epoch of radar range and Doppler measurements is nearly invisible along the bottom axis.  \label{fig5}}
\end{figure}

One of the most important results from planetary radar is highly accurate orbit determinations for near-Earth objects (asteroids and comets).  Especially when an asteroid is detected only a short time before closest approach to Earth, the orbit uncertainty from optical astrometry can be large.  A single set of radar range and Doppler measurements can reduce this uncertainty by up to three orders of magnitude (Ostro and Giorgini 2004), allowing much better predictions of potential impact hazard and more reliable recovery of the object on subsequent approaches to Earth (see Figure \ref{fig5}).

\subsection{Imaging}

\begin{figure}[!h]
\includegraphics[angle=-90,scale=0.54]{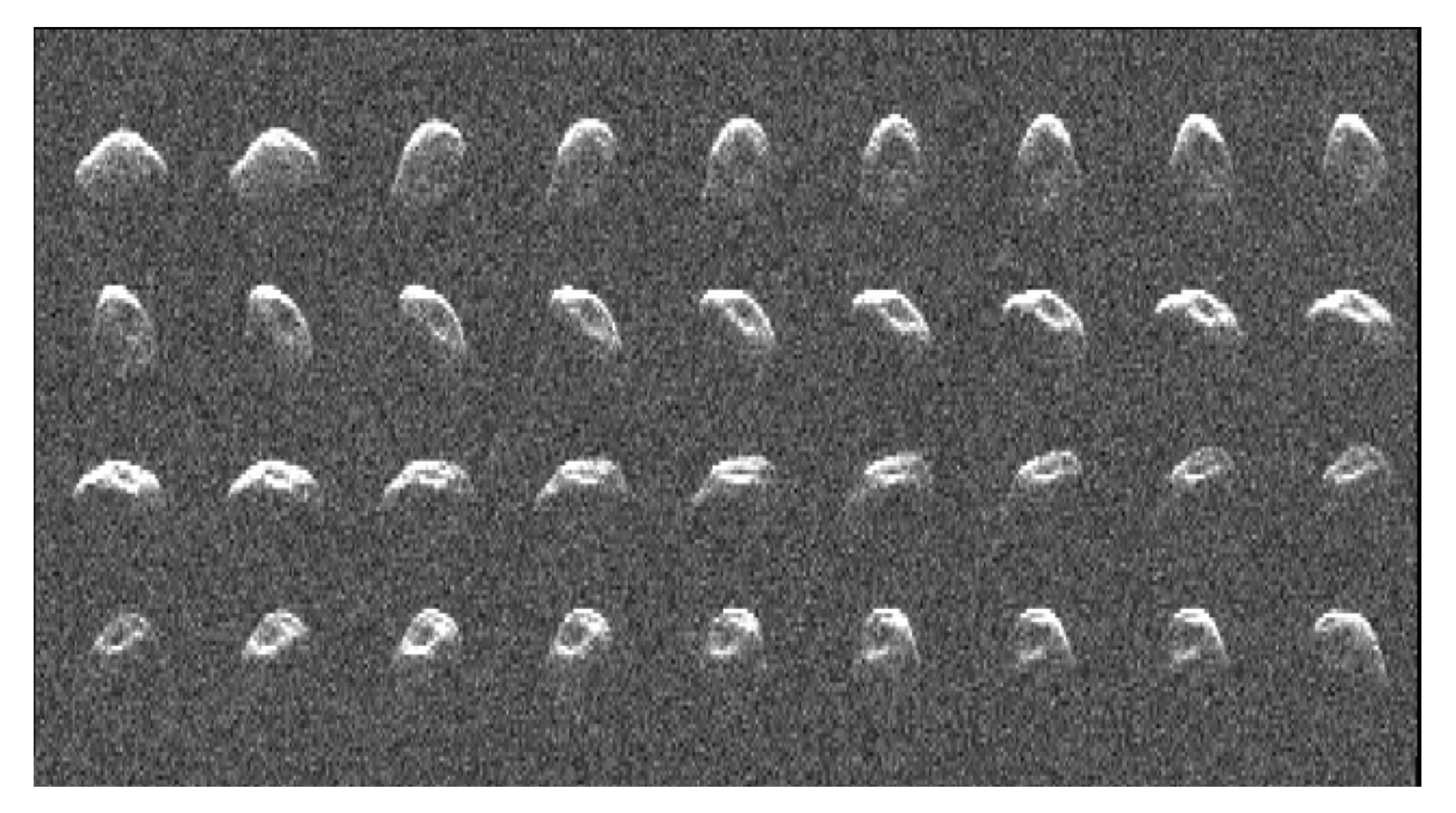}
\caption{Planetary radar images of asteroid 2010 JL33 (NASA/JPL-Caltech).  \label{fig6}}
\end{figure}

In addition to the detection and tracking of near-earth objects, radar can be used to image the shape and surface features of these objects (see Figure \ref{fig6}).  More generally, radar has been used to improve the planetary ephemeris through high accuracy range measurements to solar system bodies, and to study the surface properties of Venus, Mercury, the Galilean moons of Jupiter, and Titan, among other targets.  Imaging observations of asteroids help define the orientation of their rotation axes, an important component of orbit modeling.  All radar imaging observations would benefit from an increase in ground antenna sensitivity, even if only on the receive end of bi-static experiments.

As an example, the current Arecibo planetary radar system is able to detect Titan with an SNR approaching 100 at 2.4 GHz.  Using the full SKA in a bi-static radar observation with Arecibo would increase the SNR to $\sim$1000, and a bi-static observation with the Goldstone planetary radar at 8.4 GHz would give an SNR $\sim$10,000.  This is high enough to allow the multiple long SKA baselines to provide detailed images of the disk with a spatial resolution less than 100 km.  It also implies that Neptune's moon Triton and even Pluto could be detected with SNR $\sim$ 10-20, adequate for studies of the bulk properties of their surfaces.  

\section{Summary of SKA1-MID Requirements}

The primary requirement for SKA telemetry reception or tracking of science missions is receiver coverage of the allocated space-to-Earth downlink bands.  For deep space missions the frequency allocations are near 2300, 8400, and 32000 MHz.  The first two of these fall near the centers of receiver band 3 (1650-3050 MHz) and band 5 (4600-13800 MHz), respectively (Dewdney et al. 2013).  Band 3 of SKA1-SUR also covers the 2.3 GHz downlink band.  In some circumstances it is necessary to use UHF frequencies for data from probes, and in those cases receiver band 1 (350-1050 MHz) of SKA1-MID will be appropriate.  

\begin{figure}[!ht]
\hskip 30pt \includegraphics[angle=-0,scale=0.55]{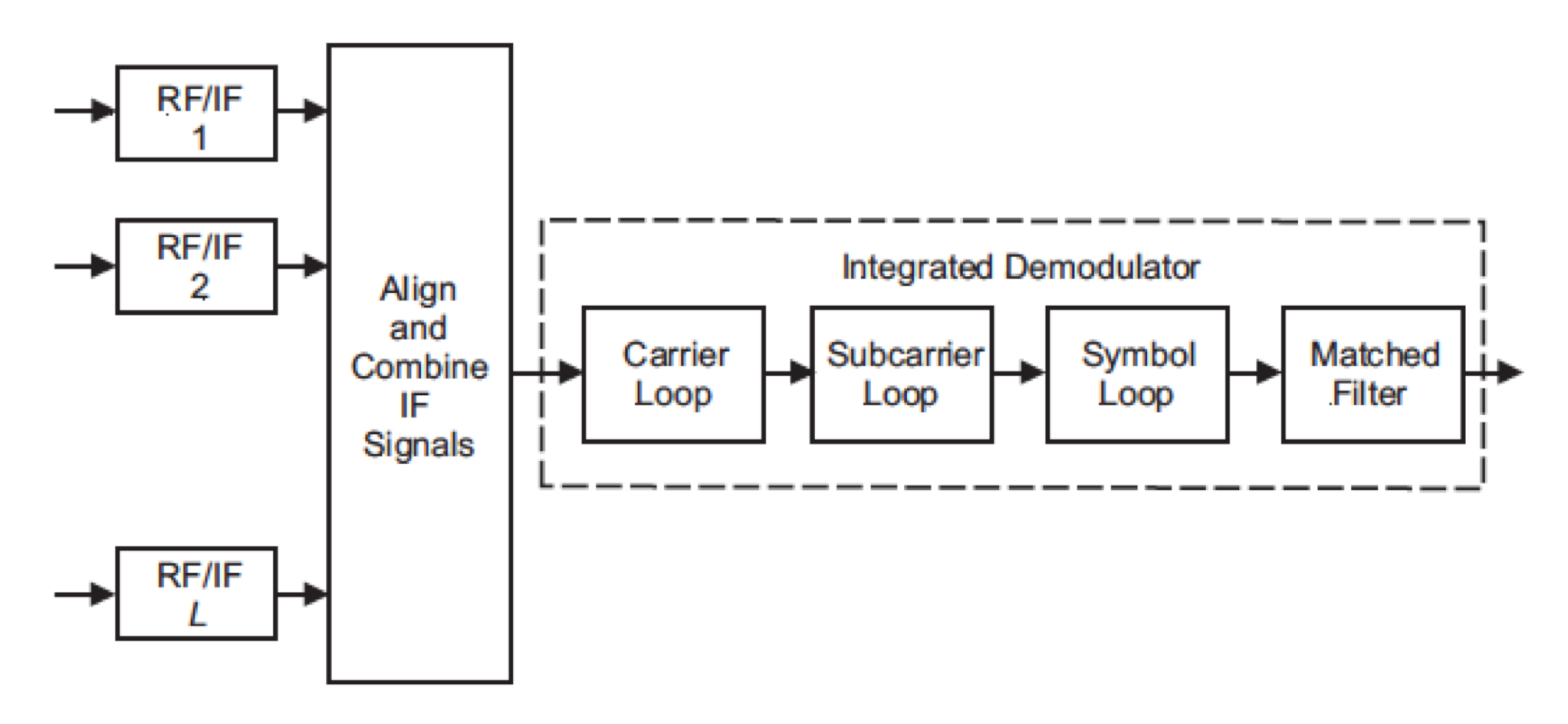}
\caption{Typical telemetry receiver architecture for an array (Rogstad et al. 2003).  Signal combining is done with the same type of beamformer used for SKA pulsar observations. The subsystem within the dashed lines would be provided by individual space missions.  \label{fig7}}
\end{figure}

Telemetry reception and spacecraft tracking require beamforming rather than cross-correlation of signals from the array antennas (see Figure \ref{fig7}).  The cost of beamforming scales linearly with the number of antennas N, compared to the N$^2$ cost scaling of cross-correlation.  In addition, spacecraft signals occupy relatively narrow bandwidths and signal processing costs scale linearly with bandwidth as well.  Only a single telemetry receiver (labeled Integrated Demodulator in Figure \ref{fig7}) is required independent of the number of antennas in the array.  JPL has developed portable telemetry receivers that are designed to be used at non-DSN sites. 

Although SKA will eventually cover the primary 2.3 GHz and 8.4 GHz downlink frequency bands, it is also interesting to speculate on the possibility of a future generation of high frequency receivers extending coverage up to the 32 GHz downlink band.  This band has a wider bandwidth allocation for space missions than the lower frequency bands, and will be more heavily used by future missions.  Since the SKA1-MID and SKA1-SUR antennas are designed to work well up to at least 20 GHz, it is quite possible that they will still have usable efficiency at 32 GHz.  Even an aperture efficiency of 10\% would provide a higher sensitivity than existing DSN 34-m antennas at 32 GHz.  Antenna blind pointing accuracy is often more of a limitation than surface accuracy in determining the high frequency limit of a radio antenna, but this is less of a concern when there is a spacecraft signal that can be used to derive real-time pointing corrections.  Thus, it is plausible that SKA could eventually receive all three of the deep space downlink frequency bands, thereby enabling tracking and telemetry reception support for future science missions to any target from any space agency.

This work was carried out at the Jet Propulsion Laboratory, California Institute of Technology, under contract with the US National Aeronautics and Space Administration.  We thank D. S. Abraham and R. A. Preston at JPL for providing Figure 1, D. W. Murphy at JPL for producing Figures 2 and 3, and J. D. Girogini at JPL for Figure 5.

\end{document}